\documentclass{aa}
\usepackage{graphicx}
\usepackage{natbib}
\usepackage{amsmath}
\usepackage{amsfonts}
\usepackage{amssymb}
\usepackage{txfonts}
\begin{document}

\title{X-ray time lags from a pivoting power law in black holes}


\author{Elmar K\"ording \inst{1} \and Heino Falcke\inst{1,2,3}}

\institute{Max-Planck-Institut f\"ur Radioastronomie,
           Auf dem H\"ugel 69, 53121 Bonn, Germany\\
           email: koerding@mpifr-bonn.mpg.de and hfalcke@mpifr-bonn.mpg.de
         \and
           Radio Observatory, ASTRON, Dwingeloo, P.O. Box 2, 7990 AA 
Dwingeloo, The Netherlands
         \and
           Dept. of Astronomy, University of Nijmegen, P.O. 
Box 9010, 6500 GL Nijmegen, The Netherlands
}

\date{08 July 2003}

\titlerunning{X-ray Variability of BHXRBs}

\abstract{
Most black hole candidate X-ray binaries show Fourier time 
lags between softer and
harder X-rays. The hard photons seem to arrive up to a few ms after the 
soft for a given Fourier frequency of the perturbation. 
The energy dependence of the time lags has a roughly logarithmic 
behavior. 
Up to now most theories fail to explain the observed magnitude and
Fourier frequency dependence of the 
lags or fail other statistical tests. 
We show that the time lags can arise from a simple pivoting power law model,
which creates the logarithmic dependence on the photon energy at once. 
A pivoting power law arises naturally from jet/synchrotron models for the
X-ray emission, but may also be applicable to corona models.
A hint to the coherence features of the light-curves can be obtained from 
the power spectral density, which can be decomposed into a few broad
Lorentzians that could arise from a couple of 
strongly damped oscillators with low  quality factors below one. 
Using small variations of the power law index for each Lorentzian separately 
the lags can be derived analytically. They show the correct 
Fourier frequency dependence of the time lags.
If one assumes variations of the power law 
index by $\pm 0.2$ the model can account for 
the observed magnitude of the time lags in Cyg~X-1. 
The model can also be applied to TeV blazars, where a pivoting power law
and hard lags have been observed directly in some cases.
As a further test we calculated the cross- and auto-correlation functions 
for our model, which also show qualitatively the observed behavior.
The auto-correlation function for higher energies has a narrower peak
than at lower energies and the cross-correlation function is
asymmetric but peaks nearly at zero.
The coherence function for the model is in agreement with the observed
data in the Fourier regime, where the model is valid. 
}

\maketitle

\section{Introduction}
The central part of active black holes seems to consist of the black hole
with an accretion disk surrounded by a hot corona (see e.g., 
\citealt{SunyaevTruemper1979}, or \citealt{HaardtMaraschi1991})
and a jet (e.g., \citealt{Spencer1979}, \citealt{MirabelRodriguez1999}, or \citealt{Fender2001}).
However, up to now the accretion flow of black holes, 
jets, their connection, and their relative prominence 
are not well understood. The most common active
black holes are active galactic nuclei (AGN) and black hole X-ray binaries
(BHXRBs). 

To constrain models and physical parameters of these objects it
is important to access all observable quantities. Besides the
spectra the variability is of high importance as it can
reveal information about the central engine and its dynamics.
Strong variability is a common phenomenom for XRBs (see e.g.,
\citealt{Klis1989}).
In BHXRBs the X-ray emission is commonly explained by an accretion disk
and a Comptonizing corona (see, e.g. \citealt{ShapiroLightmanEardley1976}, 
\citealt{SunyaevTruemper1979}, \citealt{HaardtMaraschi1991}), 
but there may also be significant contributions 
from synchrotron emission from a jet (\citealt{MarkoffFalckeFender2001} or \citealt{FalckeBiermann1999}).
The jet/synchrotron model predicts a rigid power law that can only 
vary in amplitude and in spectral index. 
Variability in Comptonization models can lead to 
a power law X-ray spectrum as well (see e.g., \citealt{KylafisKlimis1987}).
Here we will investigate whether the short term variability of active black
holes can be explained with a rigid pivoting power law model.
We will concentrate on BHXRBs as detailed light-curves are available,
but applications to AGN are as well possible.

Usually BHXRBs appear in two distinct states: the hard state 
(low flux levels accompanied with a hard power law spectrum) and
the soft-state (normally higher flux and a soft X-ray spectrum, 
see e.g., \citealt{Klis1994}).
In the hard-state a relativistic jet can usually been seen
in radio observations (e.g., \citealt{Fender2001}). We will 
focus our studies on the hard state, where the X-ray spectrum
is dominated by a power law.

A BHXRB in the hard state shows significant 
short time (0.1--100\,Hz) variability with a root mean square (rms)
around $20\% $ (see e.g., \citealt{Klis95}). 
It is therefore possible to make a detailed statistical analysis of 
the observed light curves. The light curves at
different photon energies are well correlated as the cross-correlation
function peaks nearly at unity. Furthermore, the coherence
function \citep{VaughanNowak1997} is nearly unity for a wide range
of Fourier frequencies.
However, one often  observes hard lags, e.g. the hard 
photons lag behind the soft photons up to a few milliseconds 
(e.g., see \citealt{Nolan1981}, \citealt{MiyamotoKitamoto1989}, 
\citealt{MiyamotoKimuraKitamoto1991}, or \citealt{PottschmidtWilmsNowak2000},
for a definition of phase lags see below, Eq.~\ref{eq:deflag}).
The existence of hard lags has been explained using Comptonization models.
Soft photons will be repeatedly up-scattered in a large corona, as the
harder photons need more inverse Compton processes to reach their energy
this results in hard lags. 
For studies using coronae see e.g., 
\citet{MiyamotoKimuraKitamoto1991}, \citet{NowakVaughanWilms1999}, \citet{MalzacJourdain2000}, \citet{Poutanen2002}, or \citet{BoettcherJacksonLiang2003}.
As already noted by these authors, this explanation has the problem that one
needs huge coronae and the Fourier frequency dependence of the X-ray 
time lags cannot be reproduced.

Additionally the observed auto-correlation is not 
reproduced well (see e.g., \citealt{MaccaroneCoppiPoutanen2000}).
A different approach has been made by \citet{KotovChurazovGilfanov2001},
where the authors explain the phase lags with the response
of the accretion disk to perturbations and present a short discussion of
the effects of a pivoting power law.

By the term pivoting power law we mean that the X-ray spectrum at different
times can always be described by a power law, 
which only varies in the power law index
and the overall intensity. We mostly consider the case where
the amplitude and the power law index are correlated.

The idea of a pivoting power law model
arises from recent theoretical and observational results. 
The spectrum of BHXRBs can be well described using a coupled 
jet/accretion disk model (see \citealt{MarkoffFalckeFender2001}).
Here the disk (possibly a optically thin accretion disk, 
e.g., such as ADAFs and related solutions, \citealt{NarayanYi1995}, plus a standard disk) 
is only visible
as an additional component in the UV, while the flat spectrum at radio and
optical wavelength and the power law in the X-rays is created by 
synchrotron and inverse Compton emission from the jet.
In particular, the hard X-ray power law is explained as optically thin
synchrotron emission from a single region at a few hundred 
Schwarzschild radii from the black hole.
The power law index depends on plasma parameters 
(e.g., electron temperature, adiabatic index), and may therefore
respond to changes of the jet power and the accretion rate.
As the total intensity depends on these parameters as well, 
the flux and the power law index should be correlated. 
The jet/synchrotron model therefore suggests that the X-ray emission 
behaves like a pivoting power law.
 
Within the jet/disk picture of \citet{MarkoffFalckeFender2001},
TeV Blazars like Mrk 421 or Fanaroff-Riley class I radio galaxies 
(\citealt{FanaroffRiley1974}, the unbeamed parent population
of BL Lacs within the unified scheme, \citealt{PadovaniUrry1995}) 
show many features of BHXRBs in the low/hard state, namely a 
domination of the spectral energy distribution by jet emission. 
The connection of XRBs in the hard state and jet dominated AGN is discussed in
\citet{FalckeKoerdingMarkoff2003}. 
Mrk 421, for example, shows hard lags and a positive hardness/flux correlation 
\citep{Zhang2002}.
The hardness seems to show a hysteresis effect, e.g.
the power law index seems to respond slightly after the variation of
the total intensity. 
If BHXRBs also have a power law from their jets, a similar pivoting
power law could play an important role. 
Hard lags and positive or negative hardness-flux correlations 
have also been found in Seyferts and other AGN see e.g.,
\citet{ChiangReynoldsBlaes2000} or \citet{LamerMcHardyUttley2003}.

A pivoting power law may also be applicable for Comptonization models.
Analyzing long term variability (timescales of days) of BHXRBs 
\citet{ZdziarskiGilfanovLubinski2002} suggest the existence of a
pivoting power law with a pivot point around 50 keV and explains
the behavior using Comptonization in a corona.
They find a negative correlation between flux and hardness. 
These long term variations arise probably from a different source
of variability
(e.g., the accretion rate or an other unknown parameter, 
see \citealt{HomanJonkerWijnands2002}) than the short term 
variations studied here (maybe created by magnetohydrodynamic 
instabilities, see \citealt{PsaltisNorman00}, or other unknown sources).   
Thus, it is
yet unclear if such a correlation holds for fast variations and the
true hard state.

In this paper we will analyze in a general way 
the effects of a pivoting power law model, 
where the power law index is correlated with the flux.   
We calculate the effect on the phase lags and 
the auto- and cross-correlation functions, and present a Monte Carlo simulation
of the coherence function. 
In addition to the work by \citet{KotovChurazovGilfanov2001}, 
who also discussed  the possibility that the power law index is 
directly correlated with the flux, 
we include a response time for the change of the power law index as a
function of intensity.

In Sect. 2 we describe our parameterization and model. With these
definitions we derive a general analytic solution for phase lags and
cross-correlation functions for a pivoting power law 
in Sect. 3. In Sect. 4
the analytic result is compared with a Monte Carlo simulation.
In the last two sections we discuss our model in the context of 
data from  Cygnus~X-1 and present our conclusions.

\section{Parameterization of the pivoting power law model}
As we try to calculate the time lags with an analytical approximation
it is important to parameterize our pivoting power law model
around a reference photon energy $\epsilon_0$ near the observed
energies. Let the flux $S$ of our source be a 
function of photon energy $\epsilon$ and time $t$
\begin{equation}
S(\epsilon, t) = A(t) \left(\epsilon\over \epsilon_0\right)^{-\alpha + \beta(t)} , 
\label{eq:startpl}
\end{equation}
where $\alpha$ represents the constant part of the spectral index 
while $\beta(t)$ accounts for the variations. 
The function $A(t)$ describes the flux at the reference energy $\epsilon_0$.
As we will consider the case that $A(t)$ and $\beta(t)$ are correlated, the
reference energy $\epsilon_0$ 
will not be the pivot point defined by the minimum of the rms.

If the changes in spectral index are small and we are observing
photon energies near the reference energy 
$( \ln\left(\epsilon\over \epsilon_0\right) \beta(t) \ll 1)$  
we can expand the equation
\begin{equation} 
S(\epsilon,t) = A(t) \left(\epsilon\over \epsilon_0\right)^{-\alpha} 
\left(1+\beta(t) \ln\left(\epsilon\over \epsilon_0\right)\right),
\end{equation}
and find in Fourier space, denoted by $\hat S$:
\begin{equation} 
\hat S(\epsilon,\omega) = \left(\epsilon\over \epsilon_0\right)^{-\alpha} \left( \hat A(\omega) + \widehat{\beta A}(\omega) 
 \ln\left(\epsilon\over \epsilon_0\right) \right) . 
\label{eq:ExpandS}  
\end{equation}

As we are interested in phase lags, depending on
the coherence features of $A(t)$, it is inappropriate
to use red noise for the light curve. Information on
the coherence of the light curve can be guessed from the power 
spectral density (PSD) defined as
${\rm PSD}(\omega) = \hat A^*(\omega) \hat A(\omega)$, 
where the star denotes complex conjugation.
We note that the PSD of many BHXRBs can be well described 
by a sum of a few broad Lorentzians 
\begin{equation}
{\rm PSD}(\omega) = \sum P_{\omega_i, R_i, Q_i}(\omega),
\end{equation}
with
\begin{equation}
P_{\omega_i,R_i,Q_i}(\omega) = \frac{4 R_i^2 Q_i \omega_i}{\omega_i^2 + 
4 Q_i^2 (\omega - \omega_i)^2} ,
\end{equation}
where one Lorentzian can be centered around $\omega = 0$ 
(see \citealt{Nowak2000}, \citealt{Pottschmidt2002}, or 
\citealt{BelloniPsaltis2002}).
This definition of a Lorentzian follows \citet{BelloniPsaltis2002}.
The quality factor $Q$ is a measure of the full width half maximum (FWHM)
$Q = \frac{\omega}{2 \pi {\rm FWHM}}$. The normalization factor $R$ 
describes the amplitude of the Lorentzian. It is connected to the total 
rms amplitude as
\begin{equation}
{\rm rms}_i = R_i \sqrt{ \frac{1}{2} - \frac{\tan^{-1}(-2 Q_i)}{\pi}}
\end{equation}
(see e.g., \citealt{Pottschmidt2002}).  
Lorentzians usually arise from damped oscillating systems, for
example they are used to describe the spectral shape of a laser.
The quality factor $Q$ describes how strong the oscillator is damped, 
a high $Q$ denotes a nearly undamped system with a strongly peaked
PSD, while a low value for $Q$ yields a highly damped system with
an asymmetric, weakly peaked PSD.
Usually around four Lorentzians 
with a quality factors $Q \lesssim 1$ are needed to 
fit the PSD of XRBs in the low/hard state.  
As the origin of these broad Lorentzians is
still unknown we assume that each Lorentzian is created by a strongly damped 
oscillator excited at random times. For example, these oscillators 
could be due to excitations 
at different locations on the accretion disk with a 
Fourier frequency defined
by the Keplerian rotation that may or may not be transfered into the jet. 
Other possible explanations include magnetohydrodynamic 
instabilities \citep{PsaltisNorman00} or jet precession.

To simplify the discussion we first look at only one broad Lorentzian centered
around $\omega_0$. We assume that the variability is created by a damped 
oscillator. To generate our light-curve we use a simple shot noise model 
(for shot noise models see e.g., \citealt{Terrell1972}, \citealt{LochnerSwankSzymkowiak1991} or  \citealt{NegoroKitamotoMineshige2001}). 
Let us first assume that the light curve of the BHXRB can be described 
as $F(t)$, if this oscillator has only 
been excited at $t=0$ with a unit excitation.
The overall light curve will be a superposition of many 
excitations at random times and amplitudes. If $\lambda_i$ describes the
amplitude of the excitation at the time $t_i$ we can write 
\begin{equation}
A(t) = A_{\rm DC} + \sum_i \lambda_i F(t-t_i) , \label{eq:LightCurve}
\end{equation}
where $A_{\rm DC}$ describes the constant offset of the flux.
The amplitudes $\lambda_i$ and the excitation times $t_i$ are
random variables. We choose the normalization of $F(t)$ such that
$\langle\lambda^2\rangle = 1$. Using this process we create a light curve that
has the observed PSD and the coherence properties given by the oscillator.

It is unclear whether such simple shot noise models can describe the
light-curves of XRBs (see e.g., \citealt{LochnerSwankSzymkowiak1991}). 
However, we have to disentangle the contributions
of the different broad Lorentzians to apply our model, so it is very hard to 
test a pivoting power law  model using observed light curves. 
Shot noise is therefore one of the best possibilities to create artificial
light curves available. 
Furthermore, this approach enables us to give an analytic solution for
the phase lags and the cross-correlation function.
We discuss below the effect of different
coherence properties. The main result is likely to be independent of
the shot noise assumption.

Transfered into Fourier-space we find
\begin{equation}
A(\omega) = A_{\rm DC} \delta(\omega) + \sum_i \lambda_i \hat{F}(\omega) 
{\rm e}^{{\rm i} \omega t_i } . \label{eq:expaexpr}
\end{equation}

The PSD of a complex damped oscillator is a Lorentzian, but as we are
interested in real solutions for the light curve we have to use
a linear combination of the real and the imaginary part of 
the damped complex oscillator. 
The two fundamental real solutions are the instantaneously 
excited oscillator (cosine) 
\begin{equation} 
\hat{F}_c(\omega) = \frac{1}{2} \left( {\hat{H}_{+} + \hat{H}_{-}}\right) 
\label{eq:dampedoszillator}
\end{equation}
and 
\begin{equation}
\hat{F}_s(\omega) = \frac{1}{2{\rm i}} \left({\hat{H}_{+} - \hat{H}_{-}}\right)  
\end{equation}
the sine combination, where 
$\hat{H}_\pm = \frac{2 R \sqrt{Q \omega_0}} {\omega_0 - 2 {\rm i} Q (\omega \pm \omega_0)}$
is the Fourier transform of the complex oscillator with frequency 
$\omega = \pm \omega_0$. We note that the spectral form of 
the cosine combination $F_c$ 
declines with $\omega^{-2}$ like the Lorentzian, while the sine term
drops with $\omega^{-4}$. 

Given the light curve $A(t)$ we have to choose a physical response of
the power law index $\beta(t)$.
Whatever model one uses for the X-ray emission (Comptonization or jet model)
the spectral index depends on physical properties near the accreting 
object. It is therefore likely that $\beta(t)$ will respond to changes of 
the accretion rate. As the region of emission has a characteristic size,
$\beta$ may not follow $A(t)$ directly, but may respond a bit later.
It is also possible that the emission mechanism has a response time itself
(e.g., for the jet model how fast 
is the particle acceleration mechanism responding). 
We will take this into account by introducing a response time $\tau$.
The first order approximation of $\beta$ will therefore depend 
linearly on $A(t-\tau)$. 
We assume that $\tau$ will be a small 
fraction $( a )$ of the period of the center frequency of the Lorentzian, i.e. 
$\tau = a 2 \pi / \omega_0$ with $0 \leq a <1$.  

The first order approximation form of $\beta$ and
the simplest form with the described properties is
\begin{equation} 
\beta(t) = \gamma A_{\rm AC}(t - \tau), 
\end{equation} 
where the subscript AC marks the contributions to the light curve from 
the damped oscillators. The constant part of a variable will be marked with
a subscript DC throughout this paper.
In the case of a positive hardness/flux correlation, as seen
in Mrk 421 and other Blazars, the parameter $\gamma$ is positive.
For a negative correlation, seen in long timescale variablity of
some BHXRBs, one has to use $\gamma <0$. 
For illustrative purposes we
first use a positive hardness/flux correlation ($\gamma > 0$) in our
calculations and discuss the other case in a separate subsection.

In Fourier representation one finds for $\beta$
\begin{equation}
\hat{\beta}(\omega) = \gamma {\rm e}^{{\rm i} \omega \tau} \hat{A}_{\rm AC}( \omega ) \hat{D}(\omega) ,  
\label{eq:beta}
\end{equation}
where we included an additional damping factor $D(\omega)$.
It should take into account the damping of the system creating the power law, 
i.e. if $\omega$ is too big, 
the oscillations are so fast that they average out, 
and the power law index $-\alpha + \beta(t)$ does not change anymore.

\section{Analytic results}

\subsection{Definition and energy dependence of phase lags}
The phase lag measures the phase change (corresponding to a time delay) 
between the light curves at two photon energies in phase space.
The Fourier phase lag $\phi(\omega)$ is defined as the argument
of the average cross power spectrum  
\begin{equation}
\phi(\omega) = {\rm arg }\  \langle \hat{S}(\epsilon_1,\omega)^* 
\hat{S}(\epsilon_2)\rangle
\label{eq:deflag} 
\end{equation}
see e.g., \citet{NowakVaughanWilms1999}. The Fourier phase lag 
can be translated to time lags by dividing through $\omega$. 
The time lags have a simple interpretation in the time domain, as they
measure the time difference between an outburst at two different
photon energies at a given Fourier frequency.
To calculate the phase lag from a light-curve we start from
\begin{equation}
\sin \phi  =  \frac{\Im \left[ \hat{S}^*(\epsilon_1,\omega) 
\hat{S}(\epsilon_2,\omega) \right] }
{|\hat{S}(\epsilon_1,\omega)| |\hat{S}(\epsilon_2,\omega)| }. 
\end{equation}

Using the linear approximation (Eq.~\ref{eq:expaexpr}) the phase lag can be
evaluated analytically (see also \citealt{KotovChurazovGilfanov2001}):
\begin{eqnarray}
\sin \phi & \approx & 
\frac{\Im \left[ [\hat{A}^* + (\widehat{\beta A})^* \ln\left(\epsilon_1\over \epsilon_0\right) ]
     [\hat{A} + (\widehat{\beta A}) \ln \left(\epsilon_2 \over \epsilon_0\right) ]  \right] }
{ \hat{A}^* \hat{A}} \nonumber  \\
 & \approx & \frac{\Im \left[  \hat{A}^* (\widehat{\beta A}) \right] }
{\hat{A}^* \hat{A}} \ln \frac{\epsilon_2}{\epsilon_1} . \label{eq:phaselag}
\end{eqnarray}
So, if $\hat{A}^* (\widehat{\beta A})$ has a nonzero imaginary part,
the phase lag will vary with $\ln \epsilon$ as seen in the observations of
BHXRBs. The photon energy dependence of the phase lag is therefore
independent of the response $\beta(t)$ of the power law index, 
but the overall magnitude and the pase lag dependence on Fourier 
frequency $\omega$ depends on the choice of $\beta$. 

\subsection{Fourier frequency dependence of the phase lag}
To derive the phase lag we start with Eq.~(\ref{eq:phaselag}). In
our linear approximation the only unknown component is the Fourier
transform of $\beta A$. Using the convolution theorem and the
Fourier transform of $\beta$ from Eq.~(\ref{eq:beta}) we get:
\begin{eqnarray}
\widehat{\beta A} & = & \gamma \int {\rm e}^{{\rm i} \omega' \tau} 
\hat{A}_{\rm AC}(\omega') \hat{A}(\omega - \omega') \hat{D}(\omega') {\rm d}\omega'  
\nonumber \\
& = & \gamma \int {\rm e}^{{\rm i} \omega' \tau}
\hat{A}_{\rm AC}(\omega') \hat{A}_{\rm AC}(\omega - \omega') \hat{D}(\omega')
{\rm d}\omega' \nonumber \\
& &  + \gamma A_{\rm DC} \hat{A}_{\rm AC}(\omega) \hat{D}(\omega) {\rm e}^{{\rm i} \omega \tau} .  
\label{eq:betaA}
\end{eqnarray}

For the case $\hat{D}(\omega)=1$ it is possible to give an
analytic calculation of the phase lag. The calculation is described in 
Appendix A and we just give the result for an instantaneously 
excited oscillator:

\begin{equation} \label{eq:phaseresult}
\begin{split}
\sin \phi  =  \gamma \ln \frac{\epsilon_2}{\epsilon_1} \biggl( &
A_{\rm DC} \sin(\tau \omega) \\
& +  \left.  \frac{\langle\lambda^3\rangle}{\langle\lambda^2\rangle} \Im \left({\rm e}^{{\rm i} \tau \omega} 
f_1(\omega) \right) 
\right),
\end{split}
\end{equation}
where $f(\omega)$ is given in the appendix. The $\lambda$ denote the random
excitation power of the oscillators.
It is important to note that
the calculations show that one can treat each excitation of the oscillator
separately -- the contributions of different excitations average out.

The first term in Eq.~(\ref{eq:phaseresult}) is the result of the
pivoting power law acting on the constant flux, while the second
term represents the pivoting power law acting on the pulse itself.
So the second term depends on $\frac{\langle\lambda^3\rangle}{\langle\lambda^2\rangle}$ and
will therefore vanish if the $\lambda$ are distributed symmetrically around
zero. In that case only the first term will contribute in that order of
the perturbation series, but higher orders can be nonzero.

\begin{figure}
\resizebox{8cm}{!}{\includegraphics{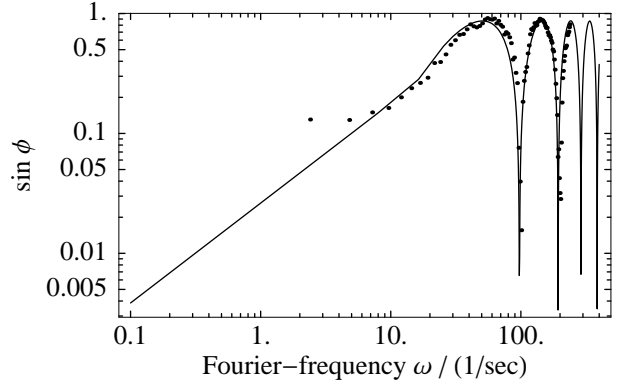}}
\caption{Phase lag dependence on the Fourier frequency. The solid line
represents the analytic approximation, the dots are the results of
a Monte Carlo simulation. The deviation of the Monte Carlo simulation 
at lower frequencies is a numerical effect.}
\label{fi:MCANA}
\end{figure}

The behavior of the phase lag is illustrated in Fig.~\ref{fi:MCANA} 
(solid line). For small $\omega \tau$ the phase lags increase linearly
with frequency. This means that the time lags are constant for
small frequencies.
Once $\omega \tau \approx \frac{\pi}{2}$, the phase lags start to 
oscillate. If $\tau$ is of the order of $\frac{2 \pi}{\omega_0}$ 
this will happen roughly at $\omega > \omega_0$.

The result that the phase lag starts to oscillate is at least partly 
due to our deterministic law of a fixed look-back time $\tau$.
A more realistic approach would be that $\tau$ is itself a random variable,
e.g., Gaussian distributed around a mean value. As the sine is
linear for $\omega \tau \ll 1$  the phase lags for low Fourier frequencies
will not change. However, for  $\omega \tau \ga 1$ the contributions
to the phase lags for different $\tau$ will average out.
The oscillation will be further reduced as  
the PSD of a real system is described
by several Lorentzians, i.e. for 
$\omega \gg \omega_0$ a second Lorentzian will dominate 
the first one. It is therefore likely that one will never
observe the oscillating part of the phase lag.

To take the probably statistical nature of $\tau$ and additional
damping of the response of the power law into account, we choose
a damping term $\hat{D}(\omega) \neq 1$, for example
$\hat{D}(\omega) = {\rm e}^{- \delta \left(\frac{\omega}{\omega_0} \right)^2}$,
which will cut off the phase lags at a given frequency. 
With an appropriately chosen
damping the phase lag will not oscillate but stay at zero for higher Fourier
frequencies.

Hence, we conclude that the phase lag in the case $\tau >0$ 
can be approximated by 
\begin{equation}
\sin \phi \sim \gamma \sin(\omega \tau) \hat{D}(\omega) 
\ln \frac{\epsilon_2}{\epsilon_1} 
\label{eq:simplelag}
\end{equation}
and is independent of the exact shape of the pulses. We find
hard lags for a positive hardness flux correlation and a positive look-back
time. 
If one changes the parameters, e.g., negative hardness flux 
correlation and positive look-back times, soft lags can be obtained.

We can now verify that our simplification made in Eq.~(\ref{eq:LightCurve})
was appropriate. We assumed that all excitations are identical and only
vary in amplitude and excitation time. In a real system each 
excitation will have a different shape.
But as the contributions to the phase lags from different excitations
average out, each pulse contributes as if there are no other excitations
(the sum in Eq.~\ref{eq:diasum} only runs over the diagonal part).
The overall phase lag will be the average of all pulses or the
phase lag of an average shaped pulse. 
In the zeroth order approximation the coherence does not play a role
for the phase lags (as long as there is a constant flux component).
Thus the shot noise assumption is not a crucial ingredient for the model.
The important assumption is only that the PSD can be decomposed into
different Lorentzians, which do not interact with each other. 

\begin{figure}
\resizebox{8cm}{!}{\includegraphics{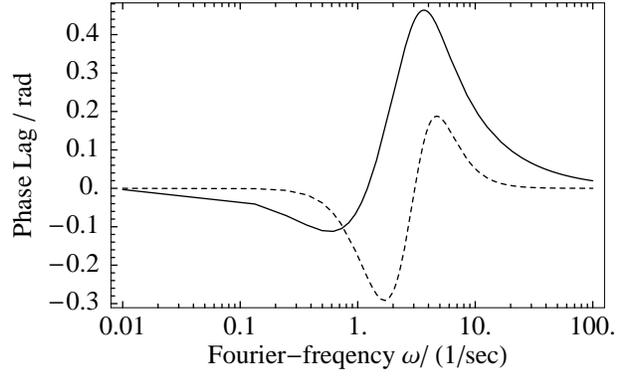}}
\caption{Difference between the sine and cosine part. The magnitude of the
sine lags (dashed) have been magnified by a factor of ten. }
\label{fi:sincos}
\end{figure}

In the case that the power law index responds instantly to a change in 
the accretion rate,  the first term of Eq.~(\ref{eq:phaseresult}) 
will vanish and
the phase lag depends to first order in $\gamma$ linearly on $\langle\lambda^3\rangle$. 
The lags would therefore vanish in the case of a symmetric 
distribution of the $\lambda$ (many systems show asymmetric excitations,
for one example see \citealt{SpruitKanbach2002}).
Furthermore, they depend strongly on the linear 
combination used for the pulse shape. 
The lags
created by the sine and cosine term are shown in Fig.~\ref{fi:sincos}.
Nevertheless, if the power law index $\Gamma$ is varying by 
$\Delta \Gamma \approx 0.2$, as will be used here later on,
the phase lags due to the pivoting power law 
will contribute significantly to the observed phase lags.  
The sign of the lag changes with frequency as mentioned by
\citet{KotovChurazovGilfanov2001}, who evaluated an instantly changing
power law for a real light curve.

\subsection{Cross-correlation Function}
The phase lags depend only on the phase of the Fourier-transform and
contain no information on the amplitude. 
To gain information also on the amplitudes we consider the
auto-correlation and cross-correlation functions.
The cross-correlation function is defined as: 
\begin{equation}
\begin{split}
C(\epsilon_1,\epsilon_2,\bar{\tau}) & =  \int \left(S(\epsilon_1,t)- \langle S(\epsilon_1)\rangle\right) \\ 
& \quad  \left(S(\epsilon_2,t+\bar{\tau})- \langle S(\epsilon_2)\rangle\right) {\rm d}t . 
\end{split}
\end{equation}
We have chosen the signs such that the cross-correlation function 
between a lower and a higher photon energy peaks to the right 
if we observe hard lags.
Expressed in Fourier-space we find:
\begin{equation}
C(\epsilon_1,\epsilon_2,\bar{\tau}) = \int 
\hat{S}_{\rm AC}(\epsilon_1,\omega)^*  \hat{S}_{\rm AC}(\epsilon_2,\omega)
{\rm e}^{ - {\rm i} \omega \bar{\tau}} {\rm d}\omega . 
\end{equation}
If we insert the expanded expression (\ref{eq:ExpandS}) 
for $\hat S(\epsilon,\omega)$  and 
only consider terms up to $O(\gamma^2)$ we find:
\begin{equation}\label{eq:ccfunkana}
\begin{split}
C(\epsilon_1,\epsilon_2,\bar{\tau}) & = \int \hat{A}^*_{\rm AC} \hat{A}_{\rm AC} {\rm d}\omega \\
& \quad + \int \left( 
\hat{A}^*\widehat{\beta A} \ln \epsilon_2  +  \widehat{\beta A}^* \hat{A} \ln \epsilon_1 
\right)_{\rm AC} 
{\rm e}^{- {\rm i} \omega \bar{\tau}}  {\rm d}\omega.
\end{split}
\end{equation}
For simplicity we have set the reference energy $\epsilon_0 = 1$.

The first component of the integrand represents the auto-correlation function
of the light curve at the reference energy:
\begin{equation}
\int \hat{A}^*_{\rm AC} \hat{A}_{\rm AC} {\rm e}^{- {\rm i} \omega \bar{\tau}}  {\rm d}\omega  = \int 
\sum_i \lambda_i^2 \hat{F}^2(\omega) {\rm e}^{- {\rm i} \omega \bar{\tau}} =: P(\bar\tau). 
\end{equation}

The second component takes a longer calculation using the results for 
$\widehat{\beta A}$ of Appendix A and is described in Appendix B.
With the function ${\cal S}(\tau,\bar \tau)$ defined in the Appendix
we find for the cross-correlation function 
\begin{equation}\label{eq:anacc}
\begin{split}
C(\epsilon_1,\epsilon_2,\bar \tau) & =  P(\bar \tau) + \gamma 
\left( \ln \epsilon_1 P(\bar \tau - \tau) 
+ \ln \epsilon_2 P(\bar \tau + \tau) \right. \\ 
& \quad + \left.\ln \epsilon_2 {\cal S}(\tau,\bar \tau) + \ln \epsilon_1 {\cal S}(\tau,-\bar \tau) \right) .  
\end{split}
\end{equation}
As described in the Appendix ${\cal S}$ peaks at $\bar \tau = \tau$ and 
decays faster than $P(\bar \tau)$.

Whether the auto-correlation function has a steeper peak for higher
photon energies $\epsilon$ depends on the parameters. 
While the terms in the first line of Eq.~(\ref{eq:anacc}) makes the auto-correlation
function broader for higher energies ($\sim \log \epsilon$) the terms in the 
second line
have the opposite effect. For small $Q$ and $\tau$ the peak is steeper 
for higher photon energies while for larger values of $\tau$ and $Q$ the
opposite effect is found. 

If the excitations of the oscillator
are symmetric, i.e. $\langle\lambda^3\rangle=0$, ${\cal S}$ vanishes, and only higher order terms
contribute to the cross-correlation function. The feature that the 
cross-correlation function can have a steeper component for higher energies 
remains in this case as seen in the Monte Carlo simulation.

\begin{figure}
\resizebox{8cm}{!}{\includegraphics{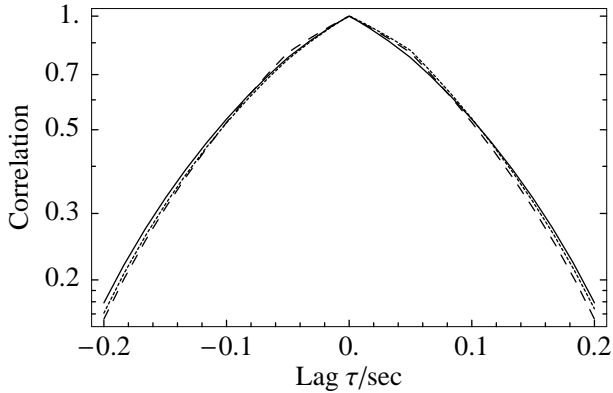}}
\caption{Numerical evaluation of the 
calculated cross-correlation function for one Lorentzian only. solid line: low energy 
auto-correlation function. dashed: higher energy ; dotted: Cross-correlation 
function }
\label{fi:NumCCFunc}
\end{figure}

A numerical evaluation of the cross correlation function with a 
damping factor of unity is shown in Fig.~(\ref{fi:NumCCFunc}). 
The calculated  auto-correlation function is, for the parameters used,
steeper for higher energy photons.
The cross-correlation function between two energies lies in between
and is slightly asymmetric. 
It is important to note that whether the auto-correlation function is
steeper for higher photon energies depends strongly on the parameters,
e.g. the look-back time $\tau$.  

If one includes a damping factor $\hat{D}(\omega)$ that damps higher
frequencies in the response of the power law index, the light
curve itself will have less power in high Fourier frequencies 
for higher energies.
Therefore the auto-correlation function will have a flatter peak
for higher energies than without a damping factor. The 
effect described above (steeper auto-correlation function for higher
energies due to the ${\cal S}$-term) and the effect of the damping factor
can cancel each other partly. 

\subsection{Negative correlation of hardness and flux}
In the previous sections we have considered a pivoting
power law model with a positive hardness flux correlation ($\gamma > 0$) and
a positive look-back time $\tau$. However \citet{ZdziarskiGilfanovLubinski2002} observed a negative
hardness flux correlation and a pivot point around 50 keV for Cyg~X-1
in long term variablity (timescale of days). 
On shorter timescales \citet{LiFengChen1999} and
\citet{FengLiChen1999} report a negative correlation as well. 
Therefore a negative correlation can also be present on the 
short timescales discussed here. 
A negative hardness/flux correlation $(\gamma <0)$ and
a positive look-back time $(\tau > 0)$ leads to soft lags.
Soft lags have for example been observed in X-ray burst oscillations 
\citep{Ford1999}. However, the observed lags are approximately 1\,rad, 
much larger than the lags discussed here. So this effect may be due to 
other mechanisms.

In order to create hard lags with a negative hardness/flux correlation
($\gamma <0$) one has to use negative look-back times $\tau$.
This means that the power law index is changed slightly 
before the flux changes.
Now the pivot point defined
by the rms is at higher photon energies as the reference 
frequency $\epsilon_0$. If $\epsilon_0 \approx 3$ keV pivot points around
50 keV can be reached.  
In this case two signs are changed in Eq.~(\ref{eq:simplelag}) 
resulting in hard lags as in the case of $\gamma >0$.
If one changes the sign of $\tau$ we have to consider that 
the analytic behavior of Eq.~(\ref{eq:betaA}) changes as well. 
The first term of the phase lag ($\sin \phi \sim \gamma \sin(\omega \tau)$)
is the same for negative and positive correlation.
However, the second term does not depend on $\Im {\rm e}^{{\rm i} \omega \tau}$ 
as before (Eq.~\ref{eq:neglagres}, see Appendix A).
We find instead
\begin{eqnarray}
\sin \phi & = & \gamma \ln \frac{\epsilon_2}{\epsilon_1} \left( 
A_{\rm DC} \sin(\tau \omega) + 
\frac{\langle\lambda^3\rangle}{\langle\lambda^2\rangle} \Im \left( f_2(\omega) \right) 
\right), 
\label{eq:neglagtext}
\end{eqnarray}
where $f_2$ is a rational function in $\omega$.
If the constant flux $A_{\rm DC}$ dominates the overall flux
the non oscillating second term will only be a minor correction.
The phase lag from a pivoting power law 
with negative look-back times $\tau$ and negative
correlation will therefore yield similar phase lags as before.
 
\begin{figure}
\resizebox{8cm}{!}{\includegraphics{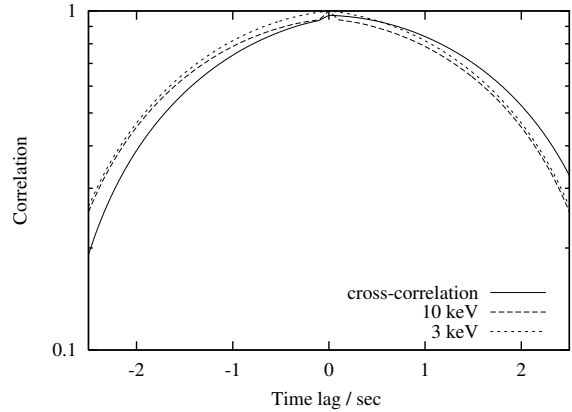}}
\caption{Monte Carlo simulation of the 
calculated cross-correlation function for a negative hardness/flux
correlation and a negative look-back time. }
\label{fi:NegCCFunc}
\end{figure}

Even though the second term of the phase lags has a different structure,
the auto-correlation function has the same analytic structure as in the
case $\tau > 0$ (Eq.~\ref{eq:anacc}). Only the function 
${\cal S}$ has to be changed. As $\gamma $ is now negative, the
first line of Eq.~(\ref{eq:anacc}) already yields a steeper auto-correlation
function for higher photon energies $\epsilon$. As the second line
depends on $\gamma$ as well, it can broaden the auto-correlation function.
Which one of these effects dominates depends on the parameters used.
Compared to the case with a positive hardness flux correlation,
the auto-correlation function has a steeper decline for small values
of $\bar \tau$ (see Fig.~\ref{fi:NegCCFunc}) and a slower one for
larger  $\bar \tau$. 

In summary, the two cases positive hardness/flux correlation and positive
look-back time and the case with a negative hardness/flux correlation and a
negative look-back time have similar hard phase lags. 
Only the parameters have to be adjusted somewhat differently.
However, while in the first case the auto-correlation function is
flatter for small lags and steepens for larger lags, the opposite effect
happens in the other case. With appropriately chosen parameters
both possibilities seem to be able to reproduce the qualitative behavior
seen in the observations of Cyg~X-1 by \citet{MaccaroneCoppiPoutanen2000}.

\section{Monte Carlo Simulations}
The analytic results have used a linear approximation for
the response of the power law. For observations near
the reference photon 
energy ${\epsilon \over \epsilon_o} \la 1$ and small responses
of the power law index $\gamma \ll 1$ this approximation will be valid.
However, as one often observes photon energies with 
${\epsilon_{\rm high} \over \epsilon_{\rm low}} \ga 10$ we created a Monte Carlo
simulation. This will also enable us to consider a system with more
than one broad Lorentzian. 
The light curve for one Lorentzian was generated as described by 
Eq.~(\ref{eq:LightCurve}). The random variables $\lambda_i$ are chosen
to be the absolute value of a Gaussian distribution and the excitation
times $t_i$ are uniformly distributed.
Each individual pulse is described by 
the sum of the cosine and sine damped oscillator ($F = F_c + F_s$).
Once the light curve at the reference photon energy is created we can
calculate light curves at other photon energies using Eq.~(\ref{eq:startpl}) 
and derive the phase lags.
The comparison between the numerical results and the analytic
calculation is shown in Fig.~\ref{fi:MCANA}. We find that 
the analytic first order approximation and the Monte Carlo simulation 
are in good agreement.
 
If one assumes that the different oscillators creating the broad 
Lorentzians do not interact with each
other one can calculate a light curve for a system described by multiple 
Lorentzians by creating light curves for each Lorentzian separately and 
then superpose these
light curves to get the overall light curve. This is used
in the application to Cyg~X-1, in Sect.~\ref{sec:application}.

Besides the phase lags we can calculate the cross- and auto-correlation
functions of the light curve. The Monte Carlo simulation reproduces
the effect seen in the analytic calculations: for the
parameters used the auto-correlation function is steeper for higher energy
photons than for the low energetic once.
In Fig.~(\ref{fi:NegCCFunc}) we show the result of the Monte Carlo simulation 
with a negative look-back time. The cross-correlation function is 
asymmetric as expected for hard lags.

\section{Applications to BHXRBs}\label{sec:application}

We are now able to apply the model to the BHXRB Cygnus~X-1.
Throughout this section we use data for Cyg~X-1 which was 
taken from \citet{NowakVaughanWilms1999} and 
\citet{KotovChurazovGilfanov2001} for the 
photon energy dependence of the phase lag.

\subsection{Photon energy dependence of the phase lags} 
\begin{figure}
\resizebox{8cm}{!}{\includegraphics{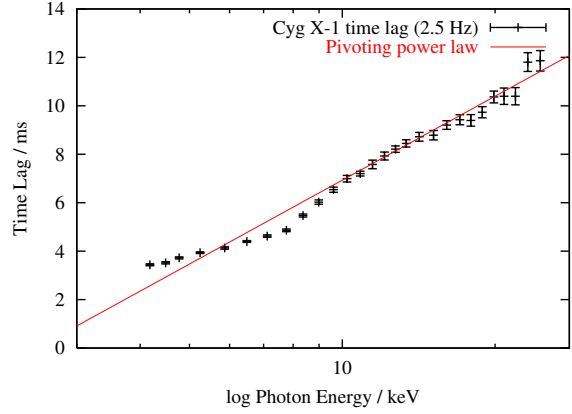}}
\caption{Time Lags versus photon energy at a Fourier frequency of
2.5\,Hz. 
The solid curve represents 
the analytic solution for the 
pivoting power law model with a reference energy of 2.0 keV. }
\label{fi:logdep}
\end{figure}

The pivoting power law model predicts that the energy dependence of
the phase lags is logarithmic (Eq.~\ref{eq:phaselag}).
In Fig.~\ref{fi:logdep} we show the phase lags measured with
a constant Fourier frequency of 2.5\,Hz (dots). The solid line
represents the pivoting power law model.  
The reference energy $\epsilon_0$ used for the fit is 2.0 keV.
For lower photon energies the accretion disk will become increasingly
important and will start to dominate over the power law component.
In this regime a simple pivoting power law will not be sufficient
to describe the behavior and deviations from the logarithmic energy
dependence of the lags are likely.  

\subsection{Fourier frequency dependence of the phase lags}
\begin{figure}
\resizebox{8cm}{!}{\includegraphics{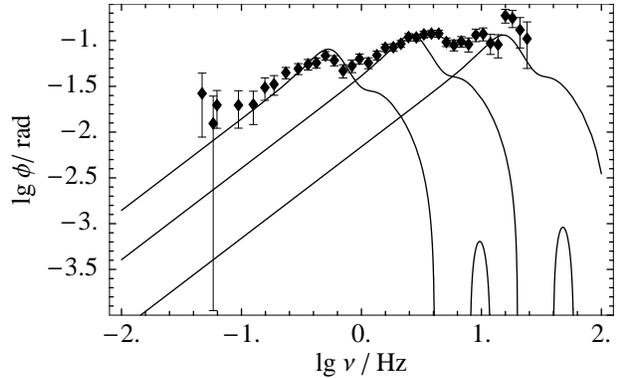}}
\caption{Illustration for the superposition of the different broad Lorentzians. In the
superposition every Lorentzian should dominate the phase lag around its center 
frequency.}
\label{fi:theosup}
\end{figure}

The PSD of Cyg~X-1 can be well fitted with four broad Lorentzians
(see e.g., \citealt{Nowak2000}, \citealt{BelloniPsaltis2002} and \citealt{Pottschmidt2002}). 
The phase lag from each Lorentzian will dominate the overall
phase lag around its center frequency. 
The idea is illustrated in
Fig.~\ref{fi:theosup} where we calculated the phase lags for
each Lorentzian separately. The look-back time $\tau$ is always the same
fraction of the center frequency of the Lorentzian, which reduces 
the number of free parameters. 
In this plot we assumed that the excitations
are similar to those of an instantaneously excited oscillator and used as damping
factor a Lorentzian with $Q=0.25$ 
(Center frequencies $f = 0.25 , 1.2, 7.3$\,Hz, $\tau = 0.03 f^{-1} $ ).

\begin{figure}
\resizebox{8cm}{!}{\includegraphics{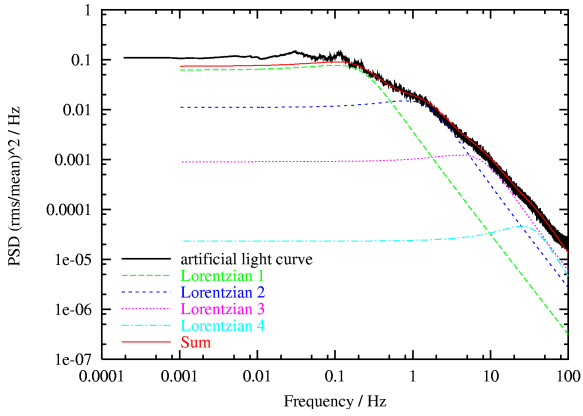}}
\caption{The PSD of the artificial light curve together with the four
broad Lorentzians. For their parameters see Table \protect\ref{tablePara}.}
\label{fi:PSD}
\end{figure}

If we assume that the different broad Lorentzians do not interact 
with each other
we can do the superposition with our Monte Carlo simulation.
First we generate a separate light curve for each Lorentzian. In this
way we have the possibility to consider 
that each broad Lorentzian has a different look-back time $\tau$, 
which we assumed to be a fixed fraction of the
period of that Lorentzian. The overall light curve is the superposition of 
the four light curves. The PSD of the artificial light curve
is shown in Fig.~\ref{fi:PSD}. The parameters of the Lorentzians are
given in Table \ref{tablePara}. The pulse of the shot noise is a
superposition of the damped sine and cosine oscillator, so its PSD
is not exactly a Lorentzian, but this only leads to small deviations 
of the overall PSD. The PSD has been used to find the overall normalization
of the shot noise.

\begin{figure}
\resizebox{8cm}{!}{\includegraphics{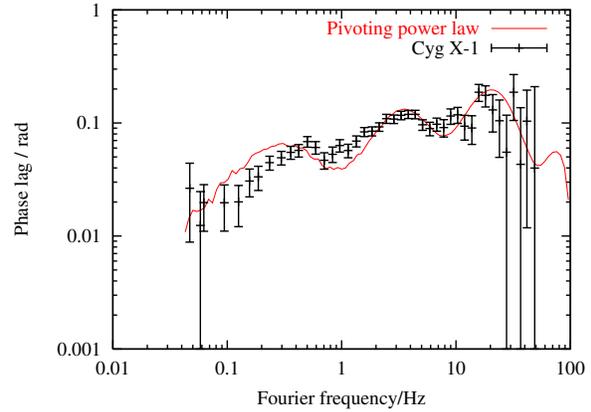}}
\caption{Phase-lag dependence 
(0--3.9\,keV to 8.2--14.1\,keV) on frequency. Four Lorentzians with 
parameters given in Table \protect\ref{tablePara} have been superposed
with our Monte Carlo Code. The data were taken from \protect\citet{NowakVaughanWilms1999}}
\label{fi:phaselag}
\end{figure}

The damping factor $D(\omega)$ used in the Monte Carlo simulation is 
a Lorentzian centered around the center-frequency of oscillator creating 
the broad Lorentzians with the half quality factor as the oscillator itself.
The constant power law index $\alpha$ has been fixed to $1.7$, and the
correlation factor $\gamma = 0.26$. The look-back time was
chosen to be $\tau = 0.04 f^{-1}$, where $f$ denotes the center frequency 
of the Lorentzian.
In Fig.~\ref{fi:phaselag} the phase lags from this Monte Carlo simulation are
plotted in comparison with the observed values from 
\citet{NowakVaughanWilms1999}.
The parameters are summarized in Table \ref{tablePara}.
The peak frequencies, where the Lorentzians contribute mostly to the rms,
are also given in this table. The peak frequency 
($f_{\rm peak} = f \sqrt{1/(4 Q^2) +1} )$ is often used for 
frequency correlations (see e.g., \citealt{Nowak2000}). 
The parameters of the Lorentzians are within the range of the 
values given by \citet{Pottschmidt2002}. The effect of the
fourth Lorentzian on the phase lags is rather small, as this
Lorentzian does not dominate the overall PSD at its center
frequency.

The used parameters are not unique. For example, a small change of 
the hardness/flux correlation parameter $\gamma$ can be compensated
by a change of the look-back time $\tau$. Furthermore, the central frequencies
of the Lorentzians needed to fit the phase lags depend on the damping factors 
and the positions of the other Lorentzians. These parameters where chosen
to give a steeper auto-correlation for higher photon energies (see below).
If one would choose different fractions of the center frequency of the
Lorentzian for the look-back time $\tau$ and allows for a different hardness/
flux correlation factor $\gamma$ for each Lorentzian better fits would
be possible. 

A similar result for the phase lags can be obtained using a negative
hardness flux correlation and a negative look-back time. One possibility
would be to choose $\gamma = -0.16$ and $\tau = -0.06 f^{-1}$ and
to center the Lorentzians given in Table \ref{tablePara} at slightly
different positions ($f = 0.15, 1.2, 6.0, 25$\,Hz). 

\begin{table}
       \caption[]{The parameters used for Cyg~X-1}
          \label{tablePara}
       \[ 
\begin{tabular}{p{0.18\linewidth}ccccc}
            \hline \hline
  Lorentzian & 1 & 2 & 3 & 4\\
             \hline
            $f_{\rm c} { \rm\  in\ Hz}$ & 0.1& 0.8 & 4.0 & 25.0 \\
            Q & 0.25 & 0.3 & 0.3 & 0.5 \\
            R & 0.22 & 0.25 & 0.16 & 0.06 \\
            $f_{\rm peak} {\rm \ in \ Hz}$ & 0.22 & 1.55 & 7.7 & 35.0 \\
             \hline
          \end{tabular}
       \]
    \end{table}
 
\subsection{Auto- and cross-correlation function}
\begin{figure}
\resizebox{8cm}{!}{\includegraphics{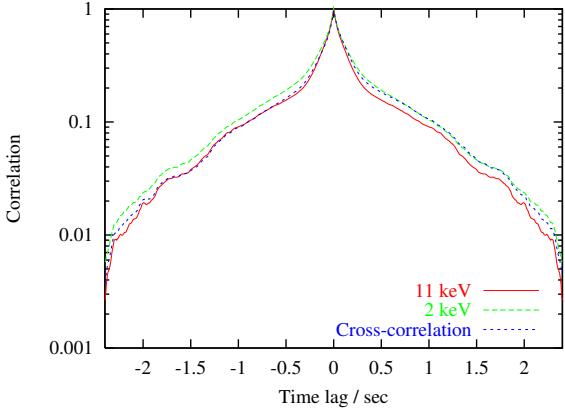}}
\caption{Auto- and cross-correlation function created by the Monte Carlo 
simulation. See the Table \protect\ref{tablePara} for the used parameters.}
\label{fi:FullMCCC}
\end{figure}

In addition to the phase lags we evaluated the auto- and cross-correlation 
function for the simulated light curves at different photon energies.
In Fig.~\ref{fi:FullMCCC} we show the auto- and cross-correlation function
for the parameters used in Fig.~\ref{fi:phaselag}, here with a positive 
hardness/flux correlation and a positive look-back time. 
With these parameters the auto-correlation function is steeper
for higher photon energies and the cross-correlation is slightly asymmetric.
The plot shows qualitatively the same behavior as 
the observed correlation shown in \citet{MaccaroneCoppiPoutanen2000}.
However, this behavior depends strongly on the parameters, such as 
the look-back time $\tau$ or even the relative prominence 
of the different Lorentzians.

\subsection{Coherence function}
\begin{figure}
\resizebox{8cm}{!}{\includegraphics{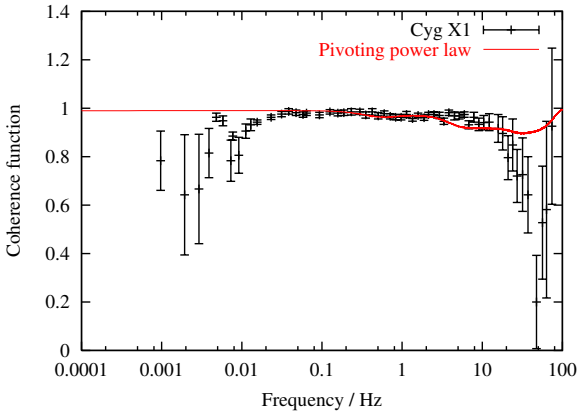}}
\caption{Coherence Function for a pivoting power law model. See the Table \protect\ref{tablePara} for the used parameters.}
\label{fi:Coherence}
\end{figure}

The coherence function measures the linear correlation of the two light
curves at different photon energies. For an introduction and a 
discussion of its properties see \citet{VaughanNowak1997}.
It is defined as
\begin{equation}
\gamma_I^2(\omega) = 
\frac{|\langle \hat{S}^*(\epsilon_1,\omega) \hat{S}(\epsilon_2,\omega) \rangle _d |^2}
{\langle |\hat{S}(\epsilon_1,\omega)|^2 \rangle _d 
\langle |\hat{S}(\epsilon_2,\omega)|^2 \rangle _d } 
\end{equation}
where $\langle . \rangle _d$ denotes an average over different realizations of
the statistical process, e.g. different light-curves.
If the coherence function is unity the light-curves at different photon
energies are connected by a linear transformation. 
To evaluate the coherence function for the pivoting power law model we
simulated 100 different 5000 second light curves with a time resolution of
5 ms. The result is shown in Fig.~\ref{fi:Coherence} together
with the measured coherence function of Cyg~X-1.
The highest frequency Lorentzian included in our calculation 
is centered around 25\,Hz. Due to our damping factors the two 
light-curves are therefore perfectly correlated at higher Fourier frequencies.
This leads to the rise of $\gamma_I^2(\omega)$ for higher Fourier frequencies
to one. The rise is therefore an  artifact of our choice of the
damping factors and probably missing higher Lorentzians.
In the low Fourier frequency domain ($\leq 0.01$\,Hz) the PSD is usually
dominated by a power law noise component \citep{Nowak2000}, which may be
responsible for the lower coherence function in this regime.
We conclude that the model is consistent with the data in the Fourier
frequency regime where the model is valid.

\subsection{Failed state transitions}
\begin{figure}
\resizebox{8cm}{!}{\includegraphics{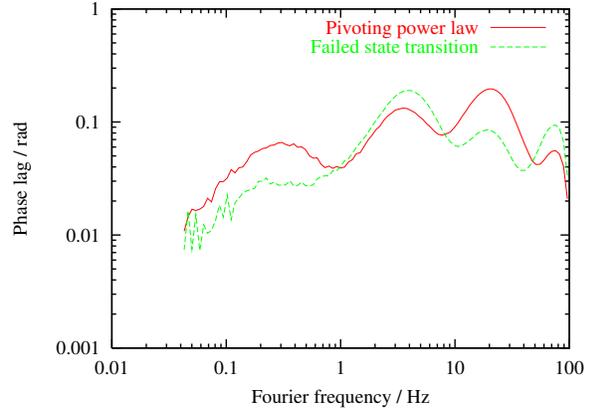}}
\caption{Comparison of the phase lags from a normal set of Lorentzians
and a 'failed state transition', where the Lorentzian two dominates. }
\label{fi:failedlag}
\end{figure}

Pottschmidt et al. (2000 \& 2002) identified some flares of Cyg~X-1
as `failed state transitions'. During these flares the X-ray spectrum
softens and that the rms amplitude decreases. The PSD is then usually
dominated only by the second Lorentzian (peak frequency $\approx 3$\,Hz) and
to some extend the third Lorentzian (peak frequency $\approx 9$\,Hz).
The phase lags during the 'failed state transitions' increases
significantly in the 3--10\,Hz regime, while it stays nearly constant
compared to a normal hard state below 1\,Hz and above 10\,Hz.
The third and second Lorentzian dominate the PSD between 3 and 10\,Hz.
This leads these authors to the conclusion that the increased phase lag
can be attributed to these Lorentzians. 
It has already been pointed
out by \citet{Nowak2000} and \citet{Pottschmidt2002} that the phase lags could be reduced due
to the superposition of the different Lorentzians. 
If one Lorentzian
starts to dominate the overall PSD the phase lags will therefore 
increase. For example in Fig.~\ref{fi:failedlag}
we reduced the strength of the first and third
Lorentzian (R = 0.1) while leaving all other parameters unchanged.
This leads to an increase of the phase lags, where the second
Lorentzian dominates, but decreases the lags further away. 
The increase and decrease of the phase lags depend on 
the other model parameters as well. While
increasing lags in the 3--10\,Hz regime are observed \citep{Pottschmidt2002},
the decrease can not be found in their plot.
However, during state transitions other model 
parameters may change. It may be that the look-back time
$\tau$ or the hardness flux correlation increases, resulting in
larger lags. To fully understand this intriguing phenomenon a more
detailed study is needed.

\section{Summary and Conclusions}
In this paper we have discussed the effect of
a variable pivoting power law in the spectrum of an astrophysical source 
on its timing behavior, particularly for the Fourier phase lags 
(see e.g., \citealt{MiyamotoKitamoto1989}),
the cross/auto-correlation and coherence function. This model is
applied to black hole X-ray binaries.
From this approach follows immediately that the
Fourier phase lag dependence on photon energy is logarithmic 
(see also \citealt{KotovChurazovGilfanov2001}), 
which is observed in Cyg~X-1.
This result is independent of the choice for the response of the power law
and the coherence features of the light curve.

To derive the Fourier frequency dependence of the phase lags, the
coherence properties of the light curve are needed. 
Hints to the coherence of XRBs can be found in the PSD, 
which can be fitted by a few broad Lorentzians
(\citealt{Nowak2000}, \citealt{Pottschmidt2002} or \citealt{BelloniPsaltis2002}). 
A Lorentzian normally arises from a damped oscillator.
Therefore, we assume that the variations of the
light-curve are generated by randomly excited damped oscillators, i.e.
we use a simple shot noise model 
(see e.g., \citealt{LochnerSwankSzymkowiak1991})
to generate our light curve.
The power law index was chosen to respond linearly to flux changes 
including a response time ($\tau$), 
i.e. the power law index responds slightly after or before 
total intensity changes.

The analytic calculations reduce the Fourier phase lags to a simple 
expression,
$\sin \phi \sim \gamma \sin (\omega \tau) \ln \frac{\epsilon_2}{\epsilon_1}$ 
for one Lorentzian, where
$\gamma$ is the flux/hardness correlation parameter and $\omega$ is the Fourier
frequency. 
This law will break down around the center frequency of the Lorentzian,
due to an included damping of the response of the power law and 
stochastic variations of the look-back time $\tau$.
The phase lag will therfore simply drop
to zero for Fourier frequencies much higher than the center frequency of
the Lorentzian.

To obtain hard lags one has to use a positive hardness/flux correlation
and a positive response time (response after the change of the flux)
or a negative hardness/flux correlation and a negative response time.
If the power law index changes by $\Delta \Gamma \approx 0.2$ around 
$\Gamma = 1.7$ and the
look-back time is of the order of $10\%$ of the period of the excitation
we can account for the observed magnitude  of the phase lags in Cyg~X-1.
Soft lags can be achieved by using a negative hardness/flux correlation
and a positive response time -- or vice versa.

The result for the phase lags of one Lorentzian is fairly independent
of the exact shape of the excitations. However, the coherence
properties of the light curve become more important if one
superposes different Lorentzians, as needed for Cyg~X-1. 
Using four Lorentzians we were able to reproduce the
observed hard lags of Cyg~X-1 \citep{NowakVaughanWilms1999} 
using parameters of the Lorentzians within the published range.
Similar hard lags are observed for BL Lacs \citep{Zhang2002}, 
where the pivoting power law model may be applicable as well.
The superposition of the Lorentzians is likely to play an important
role in the 'failed state transitions' found by 
\citet{PottschmidtWilmsNowak2000}.
During these events one Lorentzians normally dominates the overall PSD,
the effect of the superposition is reduced resulting in larger lags.

If one does not allow for a look-back time, a pivoting power law will 
nevertheless create phase lags of the order of magnitude 
of the observed values. However, these phase lags change their
sign with Fourier frequency, which is not seen in the data. 
However, when using a pivoting power law model as sometimes used to 
explain the rms behavior (see e.g., \citealt{ZdziarskiGilfanovLubinski2002}), 
one has to take these lags into account. 

Besides the phase lags we also calculated the
auto- and cross-correlation function 
(see e.g., \citealt{MaccaroneCoppiPoutanen2000}). They show the qualitative
correct behavior seen for Cyg~X-1.
It is important to note, that while the
result for the phase lags is fairly independent of the parameters and the
form of the damping factor, the auto-correlation function can change its
qualitative behavior. For example, if one uses large look-back 
times and large quality factors $Q$ the higher energy auto-correlation 
function will have a broader peak
than the one for lower energy, the opposite of what observations 
suggest for XRBs.  

The coherence function, which measures the linear correlation between
the light-curves at different photon energies \citep{VaughanNowak1997},
has been calculated and compared with observations of Cyg~X-1.
The model is in agreement with the observations, 
for Fourier frequencies where the numerical model is valid.  

Both models (the jet/synchrotron model and the disk/corona models) 
predict the existence of a pivoting power law. 
At least the jet/synchrotron model creates a rigid power law without
spectral breaks.
We have shown that such a rigid power law model is consistent 
with the data of Cyg~X-1.
A more detailed analysis using probably an original light-curve and
physical parameters for both models is needed.
We also point out, that Cyg~X-1 cannot be described by a simple power law
(see e.g., \citealt{diSalvoDoneBurderi2001}), indicating contributions from 
different emission regimes that most certainly will also complicate the
timing behavior.

We conclude that with a rather simple ansatz for 
a pivoting power law model we can explain many of the complex features in the
phase lags, cross- correlation function and coherence function seen in the
hard power law emission of XRBs.

\begin{acknowledgements}
The authors would like to thank an anonymous referee for very useful 
comments and suggestions that helped to improve the manuscript.
\end{acknowledgements}

\bibliographystyle{aa}
\bibliography{Newrefs2}

\appendix

\section{Derivation of the phase lags}
The Fourier transform of a damped oscillator is a sum of two
similar terms. To simplify the calculation let us first consider only the
first term
\begin{equation}
F_1(\omega)   =  \frac{2 R \sqrt{ Q \omega_0}} 
{\omega_0 - 2 {\rm i} Q (\omega - \omega_0)}. 
\end{equation} 
In order to evaluate the equation for the phase lags (\ref{eq:phaselag})
we have to calculate the convolution needed for $\widehat{\beta A}$ 
(Eq.~\ref{eq:betaA}):
\begin{equation}\label{eq:evalbetaA}
\begin{split}
\widehat{\beta A} & =  \gamma \int {\rm e}^{{\rm i} \omega' \tau} F_1(\omega')
F_1(\omega - \omega') \sum \lambda_i \lambda_j {\rm e}^{{ \rm i} (t_i - t_j) \omega' + {\rm i} t_j \omega} {\rm d}\omega' \\
& \quad + A_{\rm DC} F_1(\omega) \sum \lambda_j {\rm e}^{{\rm i} t_j \omega}.
\end{split}
\end{equation}
The excitation times $t_i$ are random variables, therefore  
the sum over the non-diagonal elements will yield zero as 
the random phases average out.

The integral can be solved using complex analysis.
It has two poles: one from 
$F_1(\omega')$ at $\omega' = \omega_0 - {\rm i} \frac{\omega_0}{2 Q}$
and the other from $F_1(\omega - \omega')$ near 
$\omega' = \omega - \omega_0 + {\rm i} \frac{\omega_0}{2 Q}$.
If $\tau$ is positive, we could close the integral path with a half circle for
positive imaginary parts. This path only includes the pole at $\omega' = \omega - \omega_0 + {\rm i} \frac{\omega_0}{2 Q}$.
Now we can apply the residual theorem, and find for the integral
\begin{equation}
\frac{2 \pi R^2 \omega_0}{ \omega_0 - {\rm i} Q (\omega - 2 \omega_0)} 
{\rm e}^{- \tau \frac{\omega_0}{2 Q}+ {\rm i} \tau (\omega - \omega_0)}.
\label{eq:posintegral}
\end{equation} 
Inserting this expression in Eq.~(\ref{eq:phaselag}) we get
\begin{equation}\label{eq:diasum}
\begin{split}
\sin \phi & = \gamma \ln \frac{\epsilon_2}{\epsilon_1} \biggl(  
A_{\rm DC} \sin(\tau \omega) + \\
& \frac{\Im \left ( \frac{2 \pi R^2 \omega_0}
{ \omega_0 - {\rm i} Q (\omega - 2 \omega_0)} 
{\rm e}^{- \tau \frac{\omega_0}{2 Q}+ {\rm i} \tau (\omega - \omega_0)} F^*_1(\omega) 
\sum \lambda_i^3 \right)}
{\sum \lambda_i^2 F_1^* F_1(\omega)} \Biggr). 
\end{split}
\end{equation}

The real instantaneously excited, damped oscillator is a sum of two terms like
$F_1(\omega)$. The result will therefore consist of four terms, which
can be calculated in analogy. 
We find for a positive look-back time $\tau$:
\begin{equation}\label{ap:poslagsol} 
\begin{split}
\sin \phi & =  \gamma \ln \frac{\epsilon_2}{\epsilon_1} \biggl( 
A_{\rm DC} \sin(\tau \omega) + \\
&  \qquad \frac{\langle\lambda^3\rangle}{\langle\lambda^2\rangle} \Im \left({\rm e}^{{\rm i} \tau \omega} f_1(\omega) 
\right) \biggr), 
\end{split}
\end{equation}
with 
\begin{equation}
\begin{split}
f_1(\omega) & = 
\frac{\pi R^2 \omega_0 F^*(\omega)
{\rm e}^{-\tau \frac{\omega_0}{2 Q}}}{ 2 F^* F(\omega)} \left(
\frac{{\rm e}^{- {\rm i} \tau \omega_0}} 
{ \omega_0 - {\rm i} Q (\omega - 2 \omega_0)} \right.  \\
& \quad \left. + \frac{{\rm e}^{ {\rm i} \tau \omega_0}} 
{ \omega_0 - {\rm i} Q (\omega + 2 \omega_0)}+
 \frac{({\rm e}^{ {\rm i} \tau \omega_0} +{\rm e}^{- {\rm i} \tau \omega_0} )} 
{ \omega_0 - {\rm i} Q \omega} \right).
\end{split}
\end{equation}

The other real solution for the damped oscillator is the sine 
term (see Eq.~\ref{eq:dampedoszillator}). 
For this solution one has to change the two last signs and 
include a factor $-{\rm i}$.
The only fast changing part of the second term of Eq.~(\ref{ap:poslagsol})
is ${\rm e}^{{\rm i} \tau \omega}$. Therefore, the dominant part will be 
$\sin(\tau \omega)$ as the first term.

In the second case, $\tau < 0$ we have to close 
the integral path in Eq.~(\ref{eq:evalbetaA}) with a half circle
in the negative imaginary plane. 
The integral path now includes the
pole at $\omega' = \omega_0 - {\rm i} \frac{\omega_0}{2 Q}$ and
the residual theorem yields:
\begin{equation}
\frac{2 \pi R^2 \omega_0}{ \omega_0 - {\rm i} Q (\omega - 2 \omega_0)} 
{\rm e}^{ \tau \frac{\omega_0}{2 Q}+ {\rm i} \tau \omega_0}, 
\label{eq:negintegral}
\end{equation} 
and we find for the phase lags:
\begin{equation} \label{eq:diasumneg}
\begin{split}
\sin \phi & =  \gamma \ln \frac{\epsilon_2}{\epsilon_1} \Biggl(  
A_{\rm DC} \sin(\tau \omega) +  \\
& \quad  \frac{\Im \left ( \frac{2 \pi R^2\omega_0}
{ \omega_0 - {\rm i} Q (\omega - 2 \omega_0)} 
{\rm e}^{ \tau \frac{\omega_0}{2 Q}+ {\rm i} \tau \omega_0} F_1(\omega) 
\sum \lambda_i^3 \right)}
{\sum \lambda_i^2 F_1^* F_1(\omega)} \Biggr). 
\end{split}
\end{equation}

The real instantaneously excited oscillator yields for this case:
\begin{equation}
\sin \phi  = \gamma \ln \frac{\epsilon_2}{\epsilon_1} \left( 
A_{\rm DC} \sin(\tau \omega) + 
\frac{\langle\lambda^3\rangle}{\langle\lambda^2\rangle} \Im \left( f_2(\omega) \right) 
\right), 
\label{eq:neglagres}
\end{equation}
with 
\begin{equation}
\begin{split}
f_2(\omega) & = 
\frac{\pi R^2 \omega_0   
{\rm e}^{ \tau \frac{\omega_0}{2 Q}}}{2  F^* F(\omega)} \left(
\frac{{\rm e}^{ {\rm i} \tau \omega_0} F(\omega)} 
{ \omega_0 - {\rm i} Q (\omega - 2 \omega_0)} \right.  \\
& \left.
+
 \frac{{\rm e}^{- {\rm i} \tau \omega_0} F(\omega)} 
{ \omega_0 - {\rm i} Q (\omega + 2 \omega_0)}+
 \frac{({\rm e}^{- {\rm i} \tau \omega_0} +{\rm e}^{ {\rm i} \tau \omega_0} )F(\omega)} 
{ \omega_0 - {\rm i} Q \omega} \right).
\end{split}
\end{equation}

\section{Derivation of the cross-correlation function}
In order to evaluate the second integral of the
cross-correlation function (Eq.~\ref{eq:ccfunkana})
\begin{equation}
 \int \left( 
\hat{A}^*\widehat{\beta A} \ln \epsilon_2 +  \widehat{\beta A}^* \hat{A} \ln \epsilon_1 
\right)_{\rm AC} 
{\rm e}^{- {\rm i} \omega \bar{\tau}}  {\rm d}\omega, 
\end{equation}
we start with Eq.~(\ref{eq:evalbetaA}) for $\widehat{\beta A}$.
Using this result it is possible to evaluate the integral with the same
arguments as used for $\widehat{\beta A}$.
We define three supplementary functions
\begin{gather}
\begin{split}
{\cal H}_1(\tau,\bar\tau) & = \frac{{\rm e}^{-{\rm i} \bar\tau \omega_0}}{3 \omega_0 + {\rm i} 2 Q \omega_0} +
 \frac{{\rm e}^{-{\rm i} (2 \tau - \bar\tau) \omega_0}}{3 \omega_0 + 6 {\rm i} Q \omega_0} \\
& + 2 Cos(\tau \omega_0) 
 \frac{{\rm e}^{{\rm i} (\tau - \bar\tau) \omega_0}}{3 \omega_0 - 2 {\rm i} Q \omega_0} + {\rm cc}.
\end{split}\\
\begin{split}
{\cal H}_2(\tau,\bar\tau) & = \frac{{\rm e}^{{\rm i} \tau \omega_0 - 2 {\rm i} \bar\tau \omega_0}}{3 \omega_0 + 2 {\rm i} Q \omega_0} +
\frac{{\rm e}^{{\rm i} \tau \omega_0 - 2 {\rm i} \bar\tau \omega_0}}{3 \omega_0 + 6 {\rm i} Q \omega_0}\\
& + 2 Cos(\tau \omega_0) \frac{1}{3 \omega_0 - 2 {\rm i} Q \omega_0} + {\rm cc} ,
\end{split}
\end{gather}
where cc denotes the complex conjugate, and
\begin{equation}
\begin{split}
& S(\tau, \bar \tau) = R^3 \omega_0^{3/2} Q^{-1/2} \pi^2 
{\rm e}^{-\tau \frac{\omega_0}{2 Q}}  
\left(\Theta_{\tau -\bar\tau} {\rm e}^{-(\tau - \bar\tau) \frac{\omega_0}{2 Q}}  \right. \\
& \qquad \left. {\cal H}_1(\tau,\bar \tau) +
\Theta_{\bar\tau -\tau} {\rm e}^{ 2 (\tau - \bar\tau) \frac{\omega_0}{2 Q}} {\cal H}_2(\tau,\bar \tau) \right) \sum \lambda^3,
\end{split}
\end{equation}
where $\Theta_x$ is the step function. It is defined as 
$\Theta_x = 0$ for $x<0$ and $\Theta_x = 1$ otherwise.
We find for the cross-correlation function: 
\begin{equation}\label{eq:resultcc}
\begin{split}
C(\epsilon_1,\epsilon_2,\bar \tau) & =  P(\bar \tau) + \gamma 
(\ln \epsilon_1 P(\bar \tau - \tau) + \ln \epsilon_2 P(\bar \tau + \tau)) + \\
&  \gamma \left(\ln \epsilon_2 S(\tau,\bar \tau) + \ln \epsilon_1 S(\tau,-\bar \tau) \right). 
\end{split}
\end{equation}
The first line of this equation will broaden the auto-correlation function
($\epsilon_1 = \epsilon_2$) for higher photon energies,
positive look-back times $\tau$, and positive hardness flux correlation
$\gamma > 0$.
However, ${\cal S}(\tau,\bar \tau)$ peaks at $\bar \tau = \tau$ 
and will decline faster for higher $\tau > \bar \tau$ 
than $P(\tau)$ (falls with ${\rm e}^{- \bar\tau}$) due to the factor 
${\rm e}^{ 2 (\tau - \bar\tau)}$. This leads to steeper auto-correlation functions
for higher $\epsilon$. Which effect dominates depends on the parameters
used.

For negative look-back times $\tau$ and $\gamma < 0$ we have to use
Eq.~(\ref{eq:negintegral}) for the integral of Eq.~(\ref{eq:evalbetaA}) 
resulting in a different cross-correlation function. The overall form 
will stay the same, as Eq.~(\ref{eq:negintegral}) differs from 
Eq.~(\ref{eq:posintegral}) only by the argument in the exponential function -- 
the analytic behavior (poles) is the same.
Here the first line of Eq.~(\ref{eq:resultcc}) leads to a steeper
auto-correlation function ($\gamma <0$) while the second line has the 
opposite effect.

\end{document}